\documentstyle[12pt]{article}
\newcommand{\be}{\begin{equation}}
\newcommand{\en}{\end{equation}}

\newcommand{\psin}{|\psi _n\rangle }
\newcommand{\psir}{|\psi \rangle }
\newcommand{\psip}{|\psi _p\rangle }
\newcommand{\psiz}{|\psi _z\rangle }
\newcommand{\psil}{\langle \psi }

\newcommand{\psizl}{\langle \psi _z }
\newcommand{\psipl}{\langle \psi _p}
\newcommand{\fiz}{|\varphi _z\rangle }
\newcommand{\fin}{|\varphi _n\rangle }
\newcommand{\fir}{| \varphi \rangle }

\newcommand{\finl}{\langle \varphi _n }
\newcommand{\fil}{\langle \varphi }
\newcommand{\fip}{|\varphi _p\rangle }
\newcommand{\fipl}{\langle \varphi _p}
\newcommand{\xiz}{|\xi _z\rangle }
\newcommand{\xin}{|\xi _n\rangle }

\newcommand{\xinl}{\langle \xi _n }
\newcommand{\rhoz}{|\rho _z\rangle }
\newcommand{\rhon}{|\rho _n\rangle }
\newcommand{\rhozl}{\langle \rho _z }
\newcommand{\rhonl}{\langle \rho _n }

\newcommand{\etaz}{|\eta _z\rangle }
\newcommand{\etan}{|\eta _n\rangle }

\newcommand{\etal}{\langle \eta  }

\newcommand{\Lo}{{\cal L}_0 }
\newcommand{\Li}{{\cal L}_1 }
\newcommand{\go}{\bar g_0}
\newcommand{\gi}{\bar g_1}
\newcommand{\D}{{\cal D} }

\newtheorem{defn}{Definition}
\newtheorem{rem}{Remark}
\newtheorem{lem}{Lemma}
\newtheorem{cor}{Corollary}
\newtheorem{teo}{Theorem}
\newtheorem{prop}{Proposition}

\input{amssym.def}

\begin{document}


\noindent
{\large\bf On a Resolution of the Identity in Terms of Coherent
States
}\\[10mm]

\noindent
BORIS F. SAMSONOV \\
{\it Tomsk State University, 634050 Tomsk, Russia \\
email: samsonov@phys.tsu.ru}\\[5mm]

\vspace{3mm}
\small
\noindent
{\bf Abstract.} Transformations of coherent states of the free
particle by bounded and semibounded symmetry operators are
considered. Resolution of the identity operator in terms of the
transformed states is analyzed. A generalized identity
resolution is formulated. Darboux transformation operators are
analyzed as operators defined in a Hilbert space. Coherent
states of multisoliton potentials are studied.
\vspace{5mm}

\normalsize
\noindent
{\bf 1. Introduction}

\vspace{2mm}

\noindent
The notion of coherent states is widely used in the modern
quantum mechanics and mathematical physics
\cite{kl}-\cite{Per}. The more
typical properties of coherent states are summarized in the
definition given by Klauder \cite{Klaud}. In our
interpretation
this definition looks like as follows.
\begin{defn}{\hspace{-0.5em}.} \label{def1}
Every system of states described by vectors
$\psiz $ is called the system of coherent states if the
following conditions are fulfilled{\rm :}
\newline
$({\rm i})$ $\psiz \in H_0$ where $H_0$ is a Hilbert space;
\newline
$({\rm ii})$ $z\in \D \subset \Bbb C;$
\newline
$({\rm iii})$ $\D $ is a domain endowed with a measure
$\mu (z,\bar z)$, $z,\bar z\in \D $ which is defined and finite
on a class of Borel sets of $\D $ and guaranties the following
resolution of the identity operator ${\Bbb I}$ on $H_0${\rm :}
\be \label {1}
\int _{\cal D}d\mu |\psi _z\rangle \langle \psi _z|={\Bbb I};
\en
$({\rm iv})$ $\forall z\in \D $, $\psiz $ belong to a domain of
definition of a Hamiltonian $h_0$ on $H_0$ and are
solutions to the Schr\"odinger equation
\be \label {2}
(i\partial _t-h_0)|\psi _z\rangle =0.
\en
\end{defn}

\begin{rem}{\hspace{-0.5em}.}
The integral in equation {\rm (\ref{1})} should be understood
in a
week sense. This means that if $\psin $ is an orthonormal basis
in $H_0$ then this equation is equivalent to
$\int _{\D} d \mu \langle \psi _k\psiz $ $\langle \psi _z \psin
=\delta _{nk}$. If $\xin $ is a Riesz basis in $H_0$
{\rm (}i.e. the
basis equivalent to orthonormal, see e. g. {\rm \cite{GK})}
then
formala  {\rm (\ref{1})} is equivalent to $\int _{\D} d \mu
\langle
\xi
_k\xiz \langle \xi _z \xin
=\langle \xi _k\xin $.
\end{rem}

\begin{rem}{\hspace{-0.5em}.}
In general, $\D $ is a domain in $\Bbb C^n$. In this letter we
will restrict ourselves by the case $\D =\Bbb C$. In this case
we will omit the domain of integration in the integrals.
\end{rem}

\begin{rem}{\hspace{-0.5em}.}
We introduce the property {\rm (iv)} to satisfy the condition
of
"temporal stability" formulated in {\rm \cite{Klaud}}.
\end{rem}

The condition (iii) is one of the most remarcable properties of
coherent states
widely used in mathematical physics, quantum optics, group
theory, and in
other fields of physics and mathematics. For instance, it plays
an
important role in the Berezin quantization scheme \cite{Ber},
in the analysis
of growth of holomorphic in $\D $ functions \cite{Vourd}, in a
general theory
of phase space quasiprobability distributions \cite{BM}, and in
quantum
state engineering \cite{Vo}.

In this letter we will demonstrate the insufficiency of this
definition.
In particular, we will construct a system of the vectors which
satisfy all
the conditions of the Definition \ref{def1} except for the
condition (iii). We will
show that this system satisfies a more general condition. In
this respect
we propose
a generalization of the  Definition \ref{def1}.

Our example is related with the problem of a transformation of
the 
coherent states. Let for a quantum system called initial system 
we know 
the coherent states $\psiz $ in the sense of the Definition 
\ref{def1}. We suppose 
that this system has a nontrivial symmetry operator $g_0 $ 
defined as 
usually as an operator that transforms every solution of the 
Schr\"odinger equation (\ref{2}) into another solution of the 
same 
equation. Let $\psiz $ belong to the domain of definition of 
$g_0 $. 
Consider the vectors $\fiz = g_0 \psiz $, $z\in {\cal D}$. It 
is clear 
that all the conditions of the Definition \ref{def1} except may 
be for the 
condition (iii) are 
fulfilled. Problems that may be raised in this respect are 
the following: (a) To describe the properties of $g_0 $ in 
order 
that it 
produces the coherent states in the sense of the Definition 
\ref{def1}; (b) To modify 
the condition (iii) when $\fiz $ do not satisfy the property 
(iii); (c) To 
describe the properties of $g_0 $ in order that it produces the 
coherent 
states in the sense of the modified definition. 
 
In this letter we give two examples of transformations. The 
first 
transformation is bounded and does not violate the property 
(iii). The 
states obtained with the help of the second transformation 
which is  unbounded but closed 
satisfy a more general condition then that given by the 
equation (\ref{1}). We will 
show that the integral in the  equation (\ref{1})  should be 
replaced by a 
functional defined over the set of the finite holomorphic in 
$\D $ 
functions. 
So, in this letter we will concentrate our attention  only on 
the problem 
(b) raised above. 
 
Finally we apply obtained results to coherent states of 
multisoliton potentials. For the case of the one soliton 
potential such states have been first introduced in \cite{NF}. 
 
\vspace{5mm} 
\noindent 
{\bf 2. Coherent states of the free particle} 
\vspace{5mm} 
 
\noindent 
The nonrelativistic free particle is the system very suitable 
for demonstrating  various aspects of quantum mechanics. This 
is due in particular  to the fact that the Schr\"odinger 
equation for this system has the more rich symmetry algebra. 
Moreover, the system of coherent states (in the sense of the 
Definition 1) is known for it \cite{MM}. Another important 
property of the free particle Schr\"odinger equation that 
we will 
use in this letter consists in the fact that this equation is 
the basis one for obtaining the reflectionless potentials with 
discrete energy levels disposed in the desired manner (so 
called 
multisoliton potentials, see e.g. \cite{Matv}). 
 
In this section we review briefly the well known constructions 
related to the Hilbert space of the states of the free particle 
that we will need further. 
 
Orthonormal set of solutions $\psi _n(x,t)$ of the 
Schr\"odinger equation for the free particle is well known 
\cite{Mil} and we do not cite it. We will denote by ${\rm span} 
\left\{\psi _n (x,t)\right\}$ the lineal (i.e. the space of the 
finite linear combinations) of the elements $\psi _n(x,t)$, 
$n=0,1,2,\ldots $. This lineal is an everywhere dense set in 
the space $L^2 (\Bbb R)$ of the functions square integrable on 
$\Bbb R$ with respect to the Lebesgue measure. The basis $\psi 
_n(x,t)$ has the lowering $a$ and raising $a^+$ operators, 
$a\psi _n(x,t)=\sqrt n \psi _{n-1}(x,t)$, 
$a\psi _0(x,t)=0$, 
$ a^+\psi _n(x,t)=\sqrt {n+1} \psi _{n+1}(x,t)$. The momentum 
operator $p_x$ 
 is expressed in terms of $a$ and $a^+$ as follows: 
$p_x=-i\partial _x=-(a+a^+)/2$. The free particle Hamiltomian 
is $h_0=-\partial _x^2=p_x^2$. 
 
Let us associate with the functions 
$\psi _n(x,t)=\langle x\psin $ the elements $\psin $ of an 
abstract vector space $\Lo ={\rm span}\left\{\psin \right\}$. 
Define the action of the linear raising $a$ and lowering $a^+$ 
operators on the basis elements $\psin $ by the same relations: 
$a\psin =\sqrt n|\psi _{n-1}\rangle $, $a|\psi _0\rangle =0$, 
$a^+\psin =\sqrt {n+1}|\psi _{n+1}\rangle $. Since $a$ and 
$a^+$ are supposed to be linear, their action is defined for 
every $|\psi \rangle \in \Lo $ and $a|\psi \rangle \in \Lo $, 
$a^+|\psi \rangle \in \Lo $. Moreover, $p_x\psir \in \Lo $, 
$h_0\psir \in \Lo $, $\forall \psir \in \Lo $ where 
$p_x=-(a+a^+)/2$, $h_0=p_x^2$. 
 
Let us define the scalar product $\langle \cdot |\cdot \rangle 
$ in $\Lo $, 
with the help of the coordinate 
representation $\psi (x,t)=\langle x\psir $ of the vectors 
$\psir $ and by using the ordinary Lebesgue integral. 
 
Denote by $H_0$ the completion of $\Lo $, $H_0=\bar \Lo $, with 
respect to the norm generated by this scalar product. It is 
well known that $p_x$ and $h_0$ are essentially self adjoint 
operators in $L^2(\Bbb R)$ with the well defined domains of 
definitions. In this context we will consider the closures 
$\bar p_x$ and $\bar h_0$ as the unique self adjoint extensions 
of the operators $p_x$ and $h_0$ initially defined on $\Lo $. 
The operator $\bar h_0$ is bounded from below. We will denote 
$D_{h_0}\subset H_0$ its domain of definition which is dense in 
$H_0$. In this construction the evolution parameter $t$ (time) 
is involved in every element $\psir \in H_0$. Since a one 
parametric group of evolution operators $U_t$ is continuous 
with respect to $t$ and uniquely defined by the Schr\"dinger 
equation, the derivative of $\psir $ with respect to $t$ exists 
$\forall \psir \in D_{h_0}$ \cite{Berez} and every 
$\psir \in D_{h_0}$ satisfies the Schr\"odinger equation. 
 
The operators $a$ and $a^+$ initially defined on $\Lo $ may be 
extended to a domain $D\supset \Lo $ common to both operators. 
The sum $a+a^+$, the products $aa^+$ and $a^+a$ are defined 
$\forall \psir \in D$. Moreover, 
$\langle \psi _b|a\psi _{b'}\rangle = 
\langle a^+\psi _b|\psi _{b'}\rangle $, 
$\forall |\psi _{b,b'}\rangle \in D$. 
 
The basis vectors $\psin $ are the eigenvectors of the operator 
$g_{00}=aa^+$, $g_{00}\psin =(n+1)\psin $. This operator is 
symmetric in $D$ and bounded from below. The closure $\bar 
g_{00}$ of $g_{00}$ is the unique self adjoint extension of 
this operator and it is defined $\forall \psir \in D$. 
The operator $K=\bar g_{00}^{-1}$ is a Hilbert-Schmidt 
operator and it may be chosen to equip the Hilbert space $H_0$ 
by the spaces $H_+$ and $H_-$, $H_+ \subset H_0 \subset H_-$ 
where $H_+$ is dense in $H_0$ and $H_-$ is the space of 
functionals off $H_+$. The operator $K$ may be restricted to 
$H_+$. 
The operator $K^+$ conjugate to $K$ has in this case a natural 
extension to the space $H_-$. The operator $K$ defines an 
isometry 
$H_0\rightarrow H_+$ and $K^+$ an isometry 
$H_- \rightarrow H_0$ (see e.g. \cite{Berez}). 
 
Every self adjoint in $H_0$ operator $A$ has a complete system 
of the generalized eigenvectors $|\psi _\lambda \rangle $ 
which are the functionals from $H_-$. They are defined with the 
help of the measure 
$\sigma (\lambda )=\langle e|P_\lambda e\rangle $ where 
$P_\lambda $ is the spectral function of $A$ and $|e\rangle $ 
is some element from $H_0$ as follows: 
$|\psi _\lambda \rangle = 
\frac {d|P_\lambda e\rangle }{d\sigma (\lambda )}$, 
$A|\psi _\lambda \rangle =\lambda |\psi _\lambda \rangle $ 
\cite {GSh}. The completeness of the system 
$\left\{|\psi _\lambda \rangle \right\}$ means that the Fourier 
transform $\psi (\lambda )=\langle \psi _\lambda \psir $ 
of an element $\psir \in H_0$ belongs to the space 
$L^2(d\sigma )$ of the functions square integrable with respect 
to the measure $\sigma (\lambda )$ and the Parseval equality is 
valid 
$\langle \psi \psir =\int \langle \psi |\psi _\lambda 
\rangle 
\langle \psi _\lambda \psir d\sigma (\lambda )$ for all 
$\psir \in H_0$. 
The inverse transform is written as 
$\psir =\int d\sigma (\lambda ) 
\psi (\lambda )|\psi _\lambda \rangle $. This equality should 
be understood in the week sense, i.e. 
$\forall |h\rangle \in H_+$ we have 
$\langle h\psir =\int d\sigma (\lambda )\psi (\lambda ) 
\bar h(\lambda )$. 
The latter relations may be summarized in the equation 
$$\int d\sigma (\lambda )|\psi _\lambda \rangle 
\langle \psi _\lambda |={\Bbb I}$$ 
that should be understood in the week sense. 
 
Operator $\bar p_x$ has only a continuous spectrum. Let 
$\psip $, $p\in \Bbb R$ be its generalized eigenvectors. Then 
their completeness and orthonormality may be written in the 
form 
$$\langle \psi _q\psip =\delta (q-p), 
\quad \int dp\psip \langle \psi _p|={\Bbb I}.$$ 
 
Coherent states $\psiz $ of the free particle  may be 
obtained by the action on $|\psi _0\rangle $ by the  
displacement operator \cite {Per} 
$$\psiz =\exp (za^+-\bar za)|\psi _0\rangle , \quad  
z\in \Bbb C .$$ 
These vectors are the eigenvectors of the lowering operator 
$a$, $a\psiz =z\psiz $, $ z\in \Bbb C $. The Fourier expansion 
of $\psiz $ in 
terms of the basis $\psin $ looks like as follows: 
$$\psiz =\Phi \sum _n a_nz^n\psin , \quad 
\Phi =\Phi (z,\bar z)=\exp (-z\bar z/2),\quad 
a_n=(n!)^{-1/2}.$$ 
We denote by $\bar z$ the value complex conjugate to $z$. The 
functions $\psi _z(x,t)=\langle x\psiz $ and 
$\psi _z(p,t)=\langle \psi _p\psiz $ are well known \cite{MM}. 
The vectors $\psiz $ are the coherent states in the sense of 
the Definition 1. The measure $d\mu $ in the equation (\ref{1}) 
is 
equal to 
$d\mu =dxdy/\pi $, $z=x+iy$. 
 
Since $\psip $ is the basis in $H_0$ the equation (\ref{1}) is 
equivalent to 
\be \label {dpq} 
\int d\mu \langle \psi _p\psiz \langle \psi _z|\psi _q \rangle 
=\delta (p-q). 
\en 
 
\vspace{5mm} 
\noindent 
{\bf 3. Transformation of the coherent states by symmetry 
operators} 
 
\vspace{5mm} 
 
\noindent 
Let us consider a linear symmetry operator $g_0$ initially 
defined on $\Lo $ with the help of an Hermitian matrix 
$S=\|S_{nk}\|$, $S_{nk}=\bar S_{kn}$, 
$g_0\psin =\sum _k S_{kn}|\psi _k \rangle $. We will suppose 
that every row (and column consequently) of the matrix $S$ 
contains a finite number of non zero elements. In this case the 
operator $g_0$ is symmetric in $\Lo $ and maps $\Lo \to \Lo $. 
Moreover, we will suppose that $g_0$ is bounded from below, 
positive definite, and essentially self adjoint in $H_0$ so 
that $\bar g_0=\bar g_0^+$. 
Let $D_{0} (\subset H_0)$ be the domain of definition of $\bar 
g_0$. 
 
Under these assumptions the operator $\bar g_0^{-1}$ is 
uniquely defined and bounded in $H_0$. Its domain of definition 
is the whole $H_0$. 
 
The operators $\bar g_0^{\pm 1/2}$ such that 
$\bar g_0^{\pm 1/2}\bar g_0^{\pm 1/2}=\bar g_0^{\pm 1}$ are 
uniquely defined on $H_0$ as well. The domain of definition of 
$\bar g_0^{-1/2}$ is the whole $H_0$. 
Denote $D'_{0}(\supset D_{0})$ the domain of definition of 
$\bar g_0^{1/2}$. It may be analyzed with the help of 
Friedrichs 
extension (see e.g. \cite{Smirn}) of $g_0$ up to $\bar g_0$. 
We notice that $\psin \in D_0 \subset D'_0$ 
 
\begin{lem}{\hspace{-0.5em}.} 
The systems $\left\{\rhon \right\}$, 
$\rhon =\bar g_0^{1/2}\psin $ and $\left\{\xin \right\}$, 
$\xin =\bar g_0^{-1/2}\psin $ are biorthogonal Riesz basiss 
in $H_0$ 
\end{lem} 
We will not dwell on the proof of this lemma. We note only 
that 
$$\rhonl |\rho _k\rangle =S_{nk},\ 
\xinl |\xi _k\rangle =S_{nk}^{-1}, \ 
\sum _k S_{nk} S_{kj}^{-1}=\delta _{nj},\ 
\xinl |\rho _k\rangle =\delta _{nk}.$$ 
 
 
\begin{cor}{\hspace{-0.5em}.} 
The equation {\rm (\ref{1})} may be rewritten both in terms 
 of the basis 
$\rhon $ 
$$\int d \mu \rhonl \psiz \psizl |\rho _k \rangle =S_{nk}$$ 
and in terms of the basis $\xin $ 
$$\int d \mu \xinl \psiz \psizl |\xi _k \rangle =S_{nk}^{-1}.$$ 
\end{cor} 
 
The elements $S_{nk}$ of the matrix $S$ and the elements 
$S_{nk}^{-1}$ of the matrix $S^{-1}$ may be calculated 
with the help of the generalized eigenvectors 
$|\psi _\lambda \rangle $ of the operator $\bar g_0$, 
$\bar g_0|\psi _\lambda \rangle = 
\lambda |\psi _\lambda \rangle $ 
$$S_{nk}^\gamma =\int d \sigma (\lambda ) \lambda ^\gamma 
\bar \psi _n^{(\gamma )}(\lambda ) 
\psi _k^{(\gamma )}(\lambda ), \quad \gamma =\pm 1,$$ 
$$\psi _n^{(1 )}(\lambda )=\langle \psi _\lambda \rhon , \quad 
\psi _n^{(-1 )}(\lambda )=\langle \psi _\lambda \xin $$ 
where $d\sigma (\lambda )$ is the measure that guaranties the 
spectral resolution of $\bar g_0$. 
 
It is not difficult to see that if $f(x)$ is some positive 
polynomial, $f(x)>0$, $\forall x\in \Bbb R$ and $g_0=f(p_x)$ 
then all the above assumptions imposed on $g_0$ are fulfilled. 
 
\begin{teo}{\hspace{-0.5em}.} 
If $g_0=f(p_x)$ where $f(x)$ is some 
positive polynomial in 
$x(\in \Bbb R)$ then the vectors 
$\xiz =\bar g_0^{-1/2} \psiz =\Phi \sum _n a_n z^n \xin $ 
describe coherent states in the sense of the Definition 1. 
\end{teo} 
{\bf Proof}. It is obvious that  
it is sufficient to establish the resolution of the 
identity operator. Our proof is constructive and we will only 
sketch it. 
 
Let us suppose that the measure $\mu _\xi =\mu _\xi (z,\bar z)$ 
that realizes the resolution of the identity in terms of the 
vectors $\xiz $ exists and try to find it. We will see that it 
is possible if the measure is such that 
$d\mu _\xi =\omega _\xi (x)dxdy$. 
 
To find the density $\omega _\xi (x)$ we use the generalized 
eigenvectors $\psip $ of the operator $\bar p_x$ which are the 
eigenvectors of $\bar g_0$ as well. Then using the expression 
$$\psipl \psiz =(2/\pi )^{1/4}\Phi \psi _p(z),\quad 
\psi _p(z)=\exp (-p^2+2zp-z^2/2),\quad z=x+iy,$$ 
the form (\ref{dpq}) of the formula (\ref {1}), and integrating 
with respect to the variable $y$ we arrive at the equation for 
$\omega _\xi (x)$ 
\be \label {ffm2} 
\int dx \omega _\xi (x) F_p(x)= 
(2\pi )^{-1/2}f(p)\exp (2p^2), \quad 
F_p(x)=\exp (4px-2x^2). 
\en 
It is clearly seen from this relation that the smooth function 
$\omega _\xi (x)$ is a polynomial in $x$ completely defined by 
the coefficients of the polynomial $f(p)$. This proves the 
assertion.\hfill (Q.E.D.) 
 
\begin{cor}{\hspace{-0.5em}.}  \label{cor2} 
The measure $d\mu _\xi $ is a solution to the follwing  problem 
of 
moments on the complex plane{\rm :} 
$$a_na_k\int d\mu _\xi |\Phi |^2\bar z^n z^k=S_{nk}.$$ 
\end{cor} 
This assertion follows immediately from the resolution  of the 
identity in terms of the vectors $\xiz $ and the property 
$\rhonl \xiz =\Phi a_nz^n$ (see Lemma 1). 
 
\begin{rem}{\hspace{-0.5em}.}\label {R1} 
Given the identity resolution we may construct  a holomorphic 
representation of the space $H_0$ and the operators on 
it. We do not dwell on these constructions. 
\end{rem} 
 
Let us consider the vectors 
$\rhoz =\Phi \sum _n a_n z^n \rhon $. The function 
$A_n(z)=\sum _k a_k z^k S_{nk}$ is a polynomial in $z$ since 
the sum is finite. It is not difficult to see that when 
$g_0=f(p_x)$ and $f(x)$ is a polynomial of order $2N$ then the 
number $\max _{k}S_{nk}$ have the following asymptotic 
behavior: $\max _k S_{nk}\to C n^N$ when $n\to \infty $   
where C is a constant. As a result the series 
$\sum _n a_n A_n(z)\bar z^n $ converges $\forall z\in \Bbb C$. 
This means that the vector $\rhoz $ has a finite norm and 
consequently 
$\rhoz \in H_0$. It follows that 
$\go ^{-1/2}\rhoz =\Phi \sum _n a_n z^n \psin = 
\psiz \in D'_0$ and $\go ^{1/2}\psiz =\rhoz $. 
The natural question that arises in this respect is the 
following: whether 
the states 
$\rhoz $ may be considered as coherent states. 
 
Let us suppose that the measure 
$\mu _\rho =\mu _\rho (z,\bar z)$ that realizes the resolution 
of the identity operator in terms of the vectors $\rhoz $ 
exists 
and  try to find it. Taking into account the above 
considerations we will suppose that $\mu _\rho $ is such that 
$d\mu _\rho =d\nu _\rho (x)dy$. The equation (\ref{ffm2}) takes 
in this case the form: 
\be \label {F} 
\int d\nu _\rho (x)F_p(x)=(2\pi )^{-1/2}\exp (2p^2)/f(p) 
\en 
 
We have not succeed to solve this equation in the ordinary 
functions but we have found its solution as a generalized 
function. We notice that 
$|F_p(x+iy)|\le \exp (-dx^2+by^2),\quad 2\le d\le b$. 
It follows that $F_p(x)\in S_{1/2}^{1/2}$, where the space 
$ S_{1/2}^{1/2}$ is defined as the space of entire functions 
$F$ such that $|F(x+iy)|\le \exp (-dx^2+by^2)$, 
$0<d\le b$ \cite{GS}. We may try to find $\nu _\rho $ as a 
functional off the space $ S_{1/2}^{1/2}$. 
 
It is well known \cite{GS} that a positive definite generalized 
function $\nu $ over the space $ S_{1/2}^{1/2}$ may have an 
integral representation in the space of the Fourier transforms. 
 
Let $\tilde F_p(t)=\sqrt {\pi /2}\exp (2p^2+ipt-t^2/8)$ 
be the Fourier transform of the function $F_p(x)$. We will 
understand the integral in the left hand side of the equation 
(\ref{F}) as a generalized function over the space 
$ S_{1/2}^{1/2}$ defined by the measure $\tilde \nu _\rho (t)$ 
in the space of the Fourier transforms 
\be \label {FF} 
\int d\nu _\rho (x)F_p(x)= 
\int d\tilde \nu _\rho (t)\tilde F_p(t). 
\en 
The equation (\ref{F}) results then in the equation for 
$\tilde \nu _\rho (t)$ 
\be \label{Ff} 
 \pi \int d\tilde \nu _\rho (t)\exp (-t^2/8+ipt)= 
\frac 1{f(p)}. 
\en 
 
We will give the solution to this equation for the particular 
case of the function $f(x)$ that we need further. Let $f(x)$ be 
a polynomial of order $2N$ and the zeros of $f(x)$ 
be purely imaginary. Every such a polynomial may be presented 
in the form: 
$f(x)=f_0(x)=A_0\prod _{k=1}^N (x^2+\alpha _k^2)$,  
$\alpha _k>0$. 
The value of the coefficient $A_0$ is without importance for 
our 
purpose and we 
put $A_0=1$. Then the function $1/f(x)$ may be presented in the 
form: 
$$\frac 1{f(x)}=\sum _{k=1}^N \frac {A_k}{x^2+\alpha 
_k^2},\quad 
A_k=\left[\frac {df(x)}{d(x^2)}\right]_{x^2=-\alpha _k^2}.$$ 
 
It can be seen now from (\ref{Ff}) that if 
$d\tilde \nu _\rho (t)=\tilde \omega _\rho (t)dt$ 
then for the density $\tilde \omega _\rho (t)$ we have the 
expression 
\be \label {om} 
\tilde \omega _\rho (t)=(2\pi )^{-1}\sum _{k=1}^N 
\frac {A_k}{\alpha _k}\exp (-\alpha _k|t|+t^2/8). 
\en 
 
It is necessary to note that the integral in the right hand 
side 
of 
(\ref{FF}) converges not for all functions $F(x)\in 
{S}_{1/2}^{1/2}$. 
It is easy to see that the convergence condition of this 
integral 
translates into the condition of the decreasing of the 
functions $F(x)$ when 
$|x|\to \infty $. We should take only such functions 
$F(x)\in S_{1/2}^{1/2}$ that the following inequality takes 
place: 
$|F(x)|\ge \exp (-2x^2-Ax)$ where 
$A\ge 0$ is some positive constant own to the function $F(x)$. 
Let us denote by 
$\stackrel {\circ }{S}_{1/2}^{1/2}$ 
all the functions  from $S_{1/2}^{1/2}$ satisfying this 
condition. It is apparent that $\stackrel {\circ 
}{S}_{1/2}^{1/2}$ 
is a linear space. 
 
We see hence that the integral in the left hand side of the 
equation (\ref{dpq}) should be understood as an Hermitian, 
linear on the second argument and antilinear on the first 
argument, continuous on both arguments, bounded, positive 
definite functional $\omega _\rho $ acting on the functions 
$$\psipl \rhoz =f^{1/2}(p)\psipl \psiz = 
f^{1/2}(p)\Phi \psi _p(z), \ 
\psi _p(z)=\exp (-p^2+2zp-z^2/2), \ p\in \Bbb R$$ 
as follows 
$$\omega _\rho (\psil _q\rhoz ,\rhozl \psip )=\delta (q-p).$$ 
This functional is defined with the help of the Fourier 
transform 
$\tilde F_p(t)$ of the function $F_p(x)$ defined by the 
relation 
$$\int _{-\infty }^{\infty }dy F_{pq}(z,\bar z)= 
\delta (p-q)F_p(x),\quad 
F_{pq}(z,\bar z)=|\Phi |^2\bar \psi _q(z)\psi _p(z).$$ 
 
It is not difficult to see that the function 
$$F_{a,b}(x)=\int _{-\infty}^{\infty }dy 
\langle \rho _a \rhoz \rhozl |\rho _b\rangle , \quad z=x+iy,$$ 
belongs to the space $\stackrel{\circ }{S}_{1/2}^{1/2}$ when 
$|\rho _{a,b}\rangle \in {\cal L}_\rho $. The same is true when 
we take $|\xi _{a,b}\rangle \in {\cal L}_\xi $ instead of 
$|\rho _{a,b}\rangle $. By using the biorthogonality of the 
systems $\left\{\xin \right\}$ and $\left\{\rhon \right\}$ (see 
Lemma 1) we derive that $\xinl \rhoz =\Phi a_n z^n$. It follows 
that the functional $\omega _\rho $ gives a solution to the 
following problem of moments on the complex plane: 
\be \label{omr} 
a_na_k\omega _\rho (\Phi z^n,\Phi z^k)=S_{nk}^{-1}. 
\en 
(Compare with the Corollary \ref{cor2}.) 
 
The function $\Phi $ is defined by the initial coherent states 
and it does not depend on the transformation used. This means 
that the functional $\omega _\rho $ is really defined on the 
elements $a_n z^n$ which are the basis elements in the space 
${\cal L}_\mu $ of the finite holomorphic functions. This 
functional has all the properties necessary to define a scalar 
product in ${\cal L}_\mu $. Therefore we 
may consider the completion of the lineal ${\cal L}_\mu $  with 
respect to the norm generated by this scalar product and obtain 
the Hilbert space $H_\mu =\bar {\cal L}_\mu $ of functions 
holomorphic in $\Bbb C$. (Compare with the Remark \ref{R1}.) 
 
It is now clear how we should adjust the Definition 1 to take 
in consideration the obtained results. 
\begin{defn}{\hspace{-0.5em}.} 
Let $\psin $ be such a basis in the Hilbert space  $H_0$ that 
the vectors $\psiz $ have the form{\rm :} 
$\psiz =\Phi \sum _n a_n z^n \psin $ where $a_n>0$  are some 
numbers and $\Phi =\Phi (z,\bar z)$ is some real valued 
function. By coherent states we shall mean the states described 
by the vectors $\psiz $ which satisfy the Definition 1 where 
the property {\rm (iii)} is replaced by  
{\rm (iiiþ):} in the space of the 
finite holomorphic functions there exists a functional $\omega 
$ which is Hermitian, linear in the second argument and 
antilinear in the first argument, continuous in both arguments, 
bounded, positive definite, and such that the following 
resolution of the identity operator acting in $H_0$ is valid 
$$ \omega (\psiz \Phi, \Phi \psizl |)=\Bbb I$$ 
where the equality is understood in the week sense. 
\end{defn} 
 
We summarize the above results in the following 
\begin {teo}{\hspace{-0.5em}.} 
The states 
$\rhoz =\bar g_0^{-1/2}\psiz =\sum _n a_n z^n \rhon $ 
are coherent states in the sense of the Definition 2. The 
functional $\omega =\omega _\rho $ which realizes the 
resolution of the identity in terms of the vectors $\rhoz $ 
gives the solution to the problem of moments {\rm (\ref{omr})}. 
\end{teo} 
 
\vspace{5mm} 
\noindent 
{\bf 4. Applications. Coherent states of multisoliton 
potentials } 
 
\vspace{5mm} 
 
\noindent 
It is well known (see e.g. \cite{Matv}) that the soliton 
potentials may be obtained by Darboux transformations of the 
free particle Schr\"odinger equation. The Darboux 
transformation operator, $L$, that transforms the solutions of 
the Schr\"odinger equation with zero potential to the solutions 
of the same equation with multisoliton potential is well known 
as well \cite{Matv}. 
 
Let $h_1$ be the Hamiltonian of a multisoliton potential 
having $N$ discrete spectrum levels $E_{-i}=-\alpha _{-i}^2$,
$\alpha _i>0$ 
with the eigenfunctions $\varphi _{-i}(x,t)$, $i=1,\ldots ,N$. 
If we denote $L^+$ the operator formally adjoint to $L$ then 
${\rm ker}L^+={\rm span}\left\{\varphi _{-i}\right\}$. The 
operators $L$ and $L^+$ have the remarkable factorization 
property \cite{Rev}: 
$L^+L=f(h_0)=g_0$, $LL^+=f(h_1)=g_1$ where 
$f(x)=\prod _{k=1}^N(x+\alpha _i^2)$. It should be noted that 
$h_0=p_x^2$ and $f(p_x^2)=f_0(p_x)$ where $f_0(x)$ is the 
polynomial introduced above. The functions 
$\varphi _n(x,t)=L\psi _n(x,t)$ are orthogonal to the space 
${\rm ker}L^+$. Therefor we may consider the orthogonal 
decomposition $L^2(\Bbb R)=L^2_0\oplus L^2_1$ where $L_0^2$ is 
the closure of the space ${\rm ker}L^+$ and $L^2_1$ is the 
closure of ${\rm span}\left\{\varphi _n(x,t)\right\}$, 
$n=0,1,2,\ldots $. In what follows we will not consider the 
space $L_0^2$ and concentrate our attention only on the space 
$L_1^2$. 
 
Operators $L$, $L^+$, $g_0$ and $g_1$ participate in the 
following intertwining relations: $Lg_0=g_1L$, 
$g_0L^+=L^+g_1$ which hold in the lineal 
${\rm span}\left\{\psi _n(x,t)\right\}$ and 
${\rm span}\left\{\varphi _n(x,t)\right\}$ respectively. 
Therefor if 
$g_0\psi _n(x,t)=\sum _k S_{kn}\psi _k(x,t)$ then 
$g_1\varphi _n(x,t)=\sum _k S_{kn}\varphi _k(x,t)$
(the sums are finite). 
We notice as well that $\go $ and $\gi $ are the unique self 
adjoint extensions of $g_0$ and $g_1$ respectively. 
 
Let $\psi _p(x,t)$ be the generalized eigenfunctions of 
$\bar h_0$ 
$$\bar h_0\psi _p(x,t) =p^2\psi _p(x,t),\quad 
\langle \psi _p(x,t)|\psi _q(x,t)\rangle =\delta (p-q),\quad  
p,q\in \Bbb R.$$ 
The functions 
$$\varphi _p(x,t)=N_p^{-1}L\psi _p(x,t) ,\quad  
N_p^{2}=f(p^2),\quad  
\langle \varphi _p(x,t)|\varphi _q(x,t)\rangle  
=\delta (p-q)$$ 
are the generalized eigenfunctions of $\bar h_1$, 
$\bar h_1\varphi _p(x,t) =p^2\varphi _p(x,t)$. 
 
The operator $L^+$ realizes the transformation in the inverse 
direction, $L^+\varphi _p(x,t)=N_p\psi _p(x,t)$. The functions 
$\psi _p(x,t)$ are the generalized eigenfunctions of the 
operator $\bar g_0=f(\bar h_0)$, $\bar g_0\psi _p(x,t)=f(p^2) 
\psi _p(x,t)$ and the functions $\varphi _p(x,t)$ are the 
similar ones for 
$\bar g_1=f(\bar h_1)$, $\bar g_1\varphi _p(x,t)=f(p^2) \varphi 
_p(x,t)$. 
 
In this section we will give rigorous constructions related 
with the multisoliton potentials and establish the relationship 
between the Darboux transformation operator $L$ and the 
spectral resolutions of the operators $\go $ and $\gi $. 
A polar factorization of the operator $L$ will be derived. 
Then we will introduce two kinds of coherent states  for 
multisoliton potentials. The first states are the coherent ones 
in the sense of the Definition 1 and the second states are the 
similar ones but in the sense of the Definition 2. 
 
Let us associate with the functions $\varphi _n(x,t)$ the 
vectors $\fin =L\psin $. Denote by 
${\cal L}_1={\rm span}\left\{\fin \right\}$, $n=0,1,2,\ldots $. 
The operator $L$ is supposed to be linear by definition. 
Therefor it is defined $\forall \psi \in {\cal L}_0$ and maps 
$\Lo $ onto $\Li $. Let us define for every $\fin =L\psin $, 
$n=0,1,2,\ldots $ the linear operator $L^+$ by the relation 
$L^+\fin =g_0\psin =\sum _k S_{kn}|\psi _k \rangle $ (the sum 
is finite). It is clear that $L^+$ is defined 
$\forall \fir \in \Li $ and it maps $\Li $ onto $\Lo $. 
 
Let us define the scalar product on $\Li $ by the equation 
$$\langle \varphi _a|\varphi _b\rangle _1\equiv 
\langle \psi _a|\go |\psi _b\rangle _0, \quad 
|\psi  _{a,b} \rangle \in \Lo , \quad 
|\varphi  _{a,b} \rangle =L|\psi  _{a,b} \rangle \in \Li .$$ 
Henceforth 
we label the scalar product in $H_0 $ by the subscript $0$. 
Let $H_1=\bar \Li $ be the completion of $\Li $ with respect to 
the norm generated by this scalar product. It is necessary to 
note that the set $\left\{\fin \right\}$ is a basis in $H_1$,  
$\fil _k\fin _1=S_{kn}$. We will show further that it is a 
Riesz basis and will find a basis biorthogonal with it. 
 
\begin{lem}{\hspace{-0.5em}.} 
The operator $L$ has such an extension $\bar L$ that it domain 
of definition is $D'_0$ and 
it domain of values is $H_1$. 
\end{lem} 
{\bf Proof.} Let $\psir =\sum _n c_n\psin $ be a vector from 
$D'_0$. Then 
$\langle \go ^{1/2}\psi |\go ^{1/2}\psi \rangle _0= 
\sum _{nk} \bar c_n c_k S_{nk}= c<\infty$. 
It follows that $\fil \fir _1 =c<\infty $ where 
$\fir =\sum _n c_n \fin $. So, we may put 
$\bar L\psir =\fir =\sum _n c_n \fin $, $\forall \psir \in 
D'_0$ as the definition of $\bar L$. 
It is evident that $\bar L$ is the extension of $L$ since $\bar 
L\psin =L\psin =\fin $. 
 
Consider an element $\fir \in H_1$, $\fir =\sum _n c_n \fin $, 
$\fil \fir _1=\sum _{nk} \bar c_n c_k S_{nk}=c<\infty $. 
 Consider now the 
vector $\psir \in H_0$ defined by the same Fourier 
coefficients $c_n$ with respect to the basis $\psin $, 
$\psir =\sum _n c_n\psin $. The square of the norm of  
$\go ^{1/2}\psir $ 
equal $\sum _{nk}\bar c_n c_k S_{nk}=c$ is finite. It 
follows that $\psir \in D'_0$ and by the definition of $\bar L$ 
we have $\bar L\psir =\fir $. This proves the assertion. 
\hfill (Q.E.D.) 
 
 
Let us define now for every $\fir =\bar L\psir \in H_1$, 
 such that 
$\psir \in D_0(\subset Dþ_0)$, the operator $\bar L^+$ by the 
formula 
$\bar L^+\fir =\go \psir $. Let $D_1$ be it domain of 
definition. The domain $D_1$ consists of all $\fir \in H_1$ of 
the form $\fir =\bar L\psir $ where $\psir \in D_0$. It is 
obvious that $D_1$ is dence in $H_1$. 
 
\begin{lem}{\hspace{-0.5em}.} 
$\go =\bar L^+\bar L$. 
\end{lem} 
{\bf Proof.} Since $\forall \fir \in \Lo $ we have 
$\bar L^+\bar L\fir \in \Lo $ and  
$\bar L^+\bar L\fir =g_0\fir $, the operator $\bar L^+\bar L $ 
is the extension of $g_0$ and it has the domain of definition 
equal 
$D_0$. Taking into consideration that $g_0$ has the unique 
extension $\go $ with $D_0$ as the domain of definition we get 
$\go =\bar L^+\bar L$. \hfill (Q.E.D.) 
 
\begin {lem}{\hspace{-0.5em}.} 
$\bar L^+$ is adjoint to $\bar L$ with respect to the scalar 
products $\langle \cdot |\cdot \rangle _0$ and 
$\langle \cdot |\cdot \rangle _1$ 
\end{lem} 
{\bf Proof.} The assertion follows from the chain of equalities 
$\langle \bar L^+\varphi _a|\psi _b\rangle _0= 
\langle \bar g_0\psi _a|\psi _b\rangle _0= 
\langle \psi _a|\bar g_0\psi _b\rangle _0$ $= 
\langle \varphi _a|\varphi _b\rangle _1= 
\langle \varphi _a|\bar L\psi _b\rangle _1$ 
where $|\psi _{a,b}\rangle \in D_0$, 
$|\varphi _{a,b}\rangle =\bar L|\psi _{a,b}\rangle \in D_1$. 
\hfill (Q.E.D.) 
 
\begin{lem}{\hspace{-0.5em}.}\label{lm5} 
$\bar L=\bar L^{++}$. 
\end{lem} 
{\bf Proof.} The assertion follows immediately from the 
equality $\go ^+=\go $ and Lemma 3. \hfill (Q.E.D.) 
 
\begin{cor}{\hspace{-0.5em}.} 
The operator $\bar L$ is closed. 
\end{cor}

We will take without proving the following 
\begin{prop}{\hspace{-0.5em}.} 
There exists an equipment of $H_1$, 
$H_{1+}\subset H_1\subset H_{1-}$ such that the set of 
functionals from $H_{1-}$ of the form 
$\fip =N_p^{-1}\bar L\psip $, $\psip \in H_{1-}$, 
$N_p^{2}=f(p^2)$, 
$p\in \Bbb R$, is orthonormal 
$\langle \varphi _q\fip _1=\delta (q-p)$ and complete in $H_1$. 
\end{prop} 
 
We denote by the same symbol $\bar L$ the extension of $\bar L$ 
to the space $H_{-}$. 

A similar statement has been first formulated by Krein 
\cite{Krein} for a Sturm-Liouville problem. It really means 
that the Darboux transformation operator being applied 
to a complete set of vectors of the initial system gives 
once again a complete set of vectors (in an appropriate 
space) of the transformed system. 
 
Let us denote $\gi =\bar L\bar L^+$. It follows from Lemma  
\ref{lm5} that $\gi ^+=\gi $. Moreover, since  
$\go \psip =f(p^2)\psip $, $p\in \Bbb R$, we have 
$\gi \fip =f(p^2)\fip $. This implies that if  
$\gi =f(\bar h_1)$ then $\bar h_1\fip =p^2\fip $ and the  
self-adjoint operator $\bar h_1$ is defined by the spectral  
resolution 
$$\bar h_1=\int dp p^2 \fip \fipl |$$ 
in a dense set of the space $H_1$. It is clear that $\bar h_1$ 
corresponds to the restriction of the multisoliton Hamiltonian 
to the space $L_1^2$. 
 
If we define the linear symmetric operator $g_1$ on $\Li $ by 
the formula $g_1\fin =\sum _kS_{kn}|\varphi _k\rangle $ (the 
sum is 
finite) then $\gi $ is the unique self adjoint extension of 
$g_1$. 
 
The basis $\left\{\fin \right\}$ is not orthogonal in $H_1$. We 
will construct now the basis $\left\{\etan \right\}$ 
biorthogonal to $\left\{\fin \right\}$. 
 
Since $\go $ is positive definite in $H_0$, the equation 
$\bar L^+\fir =0$, $\fir \in D_1$ has the unique solution 
$\fir =0$. It follows that $\bar L^+$ is invertible. Let us 
define $\forall \fir \in D_1$ the operator $M$ by the relation 
$M\bar L^+\fir =\fir $. 
The vector $\bar L^+\fir =\gi \psir $ belongs to the space  
$H_0$ when $\psir \in D_0$. 
Therefor the operator $M=(\bar L^+)^{-1}$ is defined on $H_0$. 
The restriction of $M$ on the lineal $\Lo $ coincides with the 
integral transformation operator introduced in 
\cite{BSjetp,BS}. 
 
Consider the vectors $\etan =M\psin $. It is clear that 
$\etal _k\fin _1=\langle \bar L\psi _k|M\psi _n\rangle _1= 
\delta _{kn}$. Therefor the system 
$\left\{\etan \right\}$ is biorthogonal to 
$\left\{\fin \right\}$. 
 
\begin {prop}{\hspace{-0.5em}.} 
Operator $U=\bar L\go ^{-1/2}$ realizes the isometric mapping 
of the domain $D'_0$ onto $D_1$. Operator 
$U^+=U^{-1}=\go ^{-1/2}\bar L^+$ realizes the inverse mapping. 
Operators $U$ and $U^+$ have the following resolutions in terms 
of the generalized eigenvectors $\psip $ and $\fip ${\rm :} 
\be \label {UU} 
U=\int dp \fip \psil _p|, \quad U^+=\int dp \psip \fil _p|. 
\en 
\end{prop} 
{\bf Proof.} Let $\psir \in D'_0$. Then $\go ^{-1/2}\psir \in 
D_0$ since $\go \go ^{-1/2}\psir =\go ^{1/2}\psir $. Therefor 
$\bar L\go ^{-1/2}\psir \in D_1$ when $\psir \in Dþ_0$. 
If $\fir \in D_1$ then 
$\bar L^+\fir =\psir \in H_0$ and 
$U^+\fir =\go ^{-1/2}\psir \in D'_0$. This means that the 
domain of values of $U^+$ is $D'_0$ and it is defined on $D_1$. 
It is easy to check that 
$U$ conserves the value of the norm 
$\psil \psir _0= 
\langle \bar L\go ^{-1/2}\psi | 
\bar L\go ^{-1/2}\psi \rangle _1$. 
The first  formula  (\ref{UU}) follows from the spectral 
resolution for $\go ^{-1/2}$. The second formula is the 
conjugation of the first. \hfill (Q.E.D.) 
 
\begin{cor}{\hspace{-0.5em}.} 
From {\rm (\ref{UU})} it follows the spectral representation 
for 
$\bar L$ and $\bar L^+$ 
$$\bar L=\int dp N_p\fip \psipl |, \quad 
\bar L^+=\int dp N_p\psip \fipl |$$ 
and the similar representation for $M$ and $M^+$ 
$$M=\int dp N_p^{-1}\fip \psipl |, \quad 
M^+=\int dp N_p^{-1}\psip \fipl |.$$ 
Operators $M$ and $M^+$ are bounded and factorize the 
operators $\go ^{-1}$ and $\gi ^{-1}$: 
$M^+M=\go ^{-1}$, $MM^+=\gi ^{-1}$. 
\end{cor} 
\begin{cor}{\hspace{-0.5em}.} 
The set $\left\{\xin \right\}$,  
$\xin =U\psin =\gi ^{1/2}\etan =\gi ^{-1/2}\fin $ is an 
orthonormal set in $H_1$ and $\left\{\etan \right\}$, 
$\left\{\fin \right\}$ are biorthogonal Riesz basiss in $H_1$. 
\end{cor} 
\begin{rem}{\hspace{-0.5em}.} 
The representation $\bar L=U\go ^{1/2}$ is a canonical 
representation of the closed operator $\bar L$ and  
$M=U\go ^{-1/2}$ is the similar representation of the bounded 
operator $M$ {\rm (}see e.g. {\rm \cite{Danf})}. These 
representations are 
called polar factorisations as well {\rm \cite{RS}}. 
\end{rem} 
 
We have seen that $\psiz \in D'_0$. Therefor the action of 
$\bar L$ on $\psiz $ is defined. Consider the vectors 
$$\fiz =\bar L\psiz =\Phi \sum _n a_n z^n \fin , \quad 
\etaz =M\psiz = \Phi \sum _n a_n z^n \etan .$$ 
 
\begin{prop}{\hspace{-0.5em}.} 
The states associated with the vectors $\etaz $ are coherent 
states in the sense of the Definition 1. The vectors $\fiz $ 
satisfy the Definition 2. 
\end{prop} 
{\bf Proof.} 
We note first of all that all the conditions of the definitions 
1 and 2 are fulfilled except may be for the property (iii) and 
(iiiþ). 
 If the measure $\mu _\eta =\mu _\eta (z,\bar z)$ that realizes 
the resolution of the identity in terms of the vectors $\etaz $ 
exists then it gives the solution to the problem of moments 
$$a_n a_k \int d\mu _\eta |\Phi |^2 z^n \bar z^k= 
\finl |\varphi _k \rangle _1= S_{nk}.$$ 
According to the Corollary 3 we have $d\mu _\eta =d\mu _\xi $. 
Similarly, if the functional $\omega  = \omega _\varphi $ that 
realizes the identity decomposition in terms of the vectors 
$\fiz $ exists then it should give the solution to the problem 
$$a_n a_k \omega _\varphi (\Phi z^n, \Phi z^k)= 
\langle \eta _n |\eta _k \rangle _1=S_{nk}^{-1}$$ 
According to the Theorem 2 we state that it is the 
functional $\omega _\rho $ that solves this problem and 
$\omega _\varphi =\omega _\rho $. This proves the assertion. 
\hfill (Q.E.D.) 
 
\vspace{5mm} 
\noindent 
{\bf 5. Conclusion} 
 
\vspace{5mm} 
 
\noindent 
In this letter we have given a rigorous mathematical meaning 
to the idea first stated in \cite{BSjetp,BS}. In short terms, 
we have formulated such a definition of coherent states that 
the states obtained by means of the Darboux transformation 
operator from coherent states of the free particle are coherent 
states of multisoliton potential. 
 The main feature of our 
definition is the existence of the resolution of the identity 
operator which has a  more general form then that ordinary used 
in 
other definitions. Nevertheless, our resolution of the identity 
permits one to construct the holomorphic representation of the 
Hilbert space of the states of a quantum system and the 
operators on it. This open the door to introduce a phase 
space  and classical observables as covariant Berezin symbols 
\cite{Ber}.  
In this way it is possible to obtain a classical 
system  the Berezin quantization of which gives the quantum 
system that we have started with. 
This program has been realized in 
\cite{jmp} for the potential $V_0(x)=\gamma x^{-2}$.  
They have shown that at classical level 
the Darboux transformation  
reduces to such a transformation of K\"ahler potential and 
Poisson bracket that the equations of motion remain unchanged. 
The similar calculations are now possible for multisoliton 
potentials. 
 
\vspace{5mm} 
\noindent 
{\bf Acknowledgments} 
\vspace{5mm} 
 
\noindent 
This work has been partially supported by RFBR grant 
\symbol{242} 9702-16279.


\begin{thebibliography}{99} 
\bibitem{kl}Klauder J. R., Skagerstam B. -S.: {\it Coherent 
States: 
Applications in Physics and Mathematical Physics}, Singapore, 
World Scientific, 1985. 

\bibitem{MM} Malkin I. A.,   Man'ko V. I.: {\it Dynamical 
Symmetries and 
Coherent States of Quantum Systems}, Nauka, Moscow (1979). 

\bibitem{Per} Perelomov A. M.:{\it Generalized Coherent States 
and Their 
Applications}, Springer, Berlin, 1986. 

\bibitem{Klaud} Klauder J. R.: Coherent States for the Hydrogen 
Atom, {\it J. Phys. A: Math. Gen.} {\bf 29} (1996), 
L293-L298. 

\bibitem{GK}Gohberg I.Ts., Krein M.G.: {\it Introduction in 
the Theory of linear Nonselfadjoint operators}, Nauka, Moscow, 
1965. 

\bibitem{Ber}Berezin F.A.: {\it The Method of Second 
Quantization}, Nauka, Moscow, 1986. 

\bibitem{Vourd}Vourdas A.: The Growth of Bargmann Functions 
and the Completeness of Sequences of Coherent States, 
{\it J. Phys. A: Math. Gen.} {\bf 30} 1997, 4867-4876. 

\bibitem{BM}Brif C., Mann A.: A general Theory of Phase-space 
Quasiprobability Distributions, {\it J. Phys. A: Math. Gen.} 
{\bf 31} 1998, L9-L17. 

\bibitem{Vo} Vourdas A.: Resolutions of the Identity in Terms 
of $SU(2)$ Coherent States and Their Use for Quantum-State 
Engineering, {\it Phys. Rev.} {\bf A 54} 1996, 4544-4552. 

\bibitem{NF}Samsonov B.F.: Coherent States of Soliton 
Potentials, {\it Phys. Atom. Nucl.} {\bf 59} 1996, 720-726. 

\bibitem{Matv} Matveev V., Salle M.: {\it Darboux 
Transformations and 
Solitons}, Springer, New York, 1991. 

\bibitem{Mil}Miller W., Jr.: {\it Symmetry and Separation of 
Variables}, Addison, London, 1977. 

\bibitem{Berez} Berezin F.A., Shubin M.A.: {\it The 
Schr\"odinger Equation}, Moscow University Press, 1983. 

\bibitem{GSh} Gel'fand I.M., Shilov G.E.: {\it The generalized 
functions}, Vol. 2, Phys.-Math. Publishing State House, Moscow, 
1958. 

\bibitem{Smirn}Smirnov V.I.: {\it Advanced Course of 
Mathematics}, Vol 5, Phys.-Math. Publishing State House, 
Moscow, 1959. 

\bibitem{GS} Gel'fand I.M., Vilenkin N. Ya.: {\it The 
generalized 
functions}, Vol. 4, Phys.-Math. Publishing State House, Moscow, 
1961. 

\bibitem{Rev} Bagrov V.G., Samsonov B.F.: Darboux 
Transformation of the Schr\"odinger Equation, 
{\it Phys. Part. Nucl.} {\bf 28} 1997, 374-397. 

\bibitem{Krein}Krein M.G.: On a Continuous Analogue of a 
Christoffel Formula from the Theory of Orthogonal Polynomials.
{\it DAN.} {\bf 113} 1957, 970-973.

\bibitem{BSjetp} Bagrov V.G., Samsonov B.F.: Coherent States of 
Anharmonic Oscillators with a Quasiequidistant spectrum, 
{\it JETP} {\bf 82} 1996, 593-599. 

\bibitem{BS} Bagrov V.G., Samsonov B.F.: Coherent States for 
Anharmonic Oscillator Hamiltonians with Equidistant and 
Quasi-equidistant Spectra, {\it J. Phys. A: Math. Gen.} 
{\bf 29} 1996, 1011-1023. 

\bibitem{Danf} Dunford N., Schwartz J.T.: {\it Linear 
Operators. Part II. Spectral Theory. Self Adjoint Operators in 
Hilbert Space}, Interscience, New York, 1963. 

\bibitem{RS}Reed M., Simon B.: {\it Methods of Modern 
Mathematical Physics. 1. Functional Analysis}, Academic, New 
York, 1972. 

\bibitem{jmp}Samsonov B.F.: Distortion of a phase space under 
the Darboux transformation, {\it J. Math. Phys.} 
{\bf 39} 1998,  967-975. 

\end{thebibliography}
\end{document}